\documentclass[aps,pre,twocolumn,groupedaddress, showkeys, a4paper]{revtex4}
\usepackage{amsmath}
\usepackage{graphicx}
\usepackage{hyperref}
\hypersetup{colorlinks=true, linkcolor=blue, citecolor=blue}

\begin{document}

\title{Exploring the limits of community detection strategies in complex networks}

\author{Rodrigo Aldecoa}
\email[]{raldecoa@ibv.csic.es}
\author{Ignacio Mar\'in}
\email[]{imarin@ibv.csic.es}
\affiliation{Instituto de Biomedicina de Valencia.
Consejo Superior de Investigaciones Científicas (IBV-CSIC)
Calle Jaime Roig 11. Valencia, Spain}

\date{\today}

\begin{abstract}
The characterization of network community structure has profound implications in several scientific areas. Therefore, testing the algorithms developed to establish the optimal division of a network into communities is a fundamental problem in the field. We performed here a highly detailed evaluation of community detection algorithms, which has two main novelties: 1) using complex closed benchmarks, which provide precise ways to assess whether the solutions generated by the algorithms are optimal; and, 2) A novel type of analysis, based on hierarchically clustering the solutions suggested by multiple community detection algorithms, which allows to easily visualize how different are those solutions. Surprise, a global parameter that evaluates the quality of a partition, confirms the power of these analyses. We show that none of the community detection algorithms tested provide consistently optimal results in all networks and that Surprise maximization, obtained by combining multiple algorithms, obtains quasi-optimal performances in these difficult benchmarks.
\end{abstract}


\maketitle

\section{Introduction}
Complex networks are widely used for modeling real-world systems in very diverse areas, such as sociology, biology and physics \cite{1,2}. It often occurs that nodes in these networks are arranged in tightly knit groups, which are called communities. Knowing the community structure of a network provides not only information about its global features, i.e., the natural groups in which it can be divided, but may also contribute to our understanding of each particular node in the network, because nodes in a given community generally share attributes or properties \cite{3}. For these reasons, characterizing which are the best strategies to establish the community structure of complex networks is a fundamental scientific problem.
 
Many community detection algorithms have been proposed so far and the best way to sort out their relative performances is by determining how they behave in standard synthetic benchmarks, consisting of complex networks whose structure is known. There are two basic types of benchmarks, which we have respectively called \textit{open} and \textit{closed} \cite{4,5,6}. Open benchmarks use networks with a community structure defined \textit{a priori}, which is progressively degraded by randomly rewiring links in such a way that the number of connections among nodes in different communities increases and the network evolves toward an unknown, ``open-ended'' structure \cite{5,6,7,8,9,10,11}. In open benchmarks, the performance of an algorithm can be measured by comparing the partitions that it obtains with the known, initial community structure, being increasingly difficult to recover that structure as the rewiring progresses. The first commonly used open benchmark was developed by Girvan and Newman (GN benchmark) \cite{12}. It is based on a network with 128 nodes, each with an average number of 16 links, split into four equal-sized communities. It is now however known that the GN benchmark is too simple. Most algorithms are able to provide good results when confronted with it \cite{7,8}. Also, the fact that all communities are identical in size makes some algorithms that favor erroneous structures (e.g., those unable to detect communities that are small relative to the size of the network \cite{6,7,8,13,14,15}) to perform artificially well in this benchmark. These results indicated the need to develop more complex benchmarks. A new type was suggested by Lancichinetti, Fortunato and Radicchi (LFR benchmarks), which has obvious advantages over the GN benchmark \cite{16}. In the GN networks, node degrees follow a Poisson distribution. However, in many real networks the degree distribution displays a fat tail, with a few highly connected nodes and the rest barely linked. This suggests that its distribution may be modeled according to a power law. In the LFR benchmarks, both the degrees of the nodes and the community sizes in the initial networks can be adjusted to follow power laws, with exponents chosen by the user. In this way, realistic networks with many communities can be built. LFR benchmarks are much more difficult than GN benchmarks, with many algorithms performing poorly in them \cite{6,8,9,10,11}. Notwithstanding these advantages, the parameters commonly used in the LFR benchmarks generate networks where all communities have similar sizes \cite{4,5,6,8}. This led to the proposal of a third type of benchmark, based on Relaxed Caveman (RC) structures \cite{17}. In this type of benchmarks, the initial networks are formed by a set of isolated cliques, each one corresponding to a community, which are then progressively interconnected by rewiring links. The possibility of selecting the size of each initial clique makes the RC benchmarks ideal for building distributions of community sizes with a high variance, which constitute a very stern test for most algorithms \cite{4,5,6}.

In open benchmarks, when the original structure is largely degraded -- and especially if the networks used in the benchmark are large and have a complex community structure -- it generally happens that all algorithms suggest partitions different from the initial one. However, this can be due to two very different reasons: either the algorithms are not performing well or all/some of them indeed are optimally recovering the community structure present in the network, but that structure does not anymore correspond to the original one. The lack of a way to discriminate between these two potential causes is a limitation of all open benchmarks. To overcome this problem, we recently proposed a different type of benchmark, which we called \textit{closed} \cite{4,5}. Closed benchmarks also start with a network with known community structure.  However, the rewiring of the links is not random, as in open benchmarks. It is instead guided from the initial network toward a second, final network, which has exactly the same community structure that the initial one, but with the nodes randomly reassigned among communities. The rewiring process in these benchmarks is called \textit{Conversion} (C), and ranges from 0 \% to 100 \%. When C = 50 \%, half of the links that must be modified in the transition from the initial to the final networks have been already rewired and C = 100 \% indicates that the final structure has been obtained. Further details about closed benchmarks can be found in one of our recent papers, in which we extensively described its behavior and how they can be applied to real cases \cite{5}

The main advantage of the closed benchmarks is that it is possible to obtain quantitative information regarding whether a given partition is optimal or not, using analyses based on parameter called Variation of Information (VI \cite{18}). By definition, VI(X,Y) = H(X) + H(Y) - 2I(X,Y), where X and Y are the two partitions to be compared, H(X) and H(Y) are the entropies of each partition and I(X,Y) is the mutual information between X and Y. The logic behind VI is to provide a distance measure indicating how different are two partitions once it is taken into account not only their structures but also the common features present in both of them. A very important reason for using VI is that it is a metric \cite{18}. In the context of the closed benchmarks, this means that VI satisfies the triangle inequality: VI$_{IE}$ $+$ VI$_{EF}$ $\geq$ VI$_{IF}$, where: 1) VI$_{IE}$ is the variation of information for the comparison between the original community structure known to be present in the initial network (I) and the one deduced for an intermediate network (E), generated at a certain point of the conversion process; 2) VI$_{EF}$ is obtained comparing that intermediate structure and the community structure of the final network (F), which is also known; and, 3) VI$_{IF}$ is obtained when the initial and final structures are compared. An algorithm that performs optimally during the whole conversion process should generate solutions satisfying the equality VI$_{IE}$ + VI$_{EF}$ $=$ VI$_{IF}$ -- where E is in this context the partition proposed by the algorithm -- while deviations from this equality, which can be summarized with the value VI$_\delta$ = VI$_{IF}$ - (VI$_{IE}$ + VI$_{EF}$), indicate suboptimal performance \cite{4,5}. Therefore, how optimal is the behavior of an algorithm can be quantified. A second advantage of the closed benchmarks is that the identical community structure in the original and final networks implies that the solutions provided by an algorithm must be symmetrical along the conversion of one into the other. For example, at C = 50 \%, a correct partition must be equally similar to both the initial and final networks. Finally, it is also significant to point out that closed benchmarks are very versatile, given that any network, for example those traditionally used in open GN, LFR or RC benchmarks, can be also analyzed in a closed configuration.

All the analyses described so far, in both open and closed benchmarks, require the community structure to be known a priori. Additional useful information may be obtained by evaluating the results of the different algorithms with measures able to establish the quality of a partition by criteria that are independent of knowing the structures originally present in the networks. In the past, one such global measure of partition quality, called modularity \cite{19}, was extensively used. However, multiple works have shown that modularity-based evaluations are often erroneous \cite{4,6,13,14,15}. In recent studies, we introduced a new global measure, called Surprise (S), which has an excellent behavior in all networks \cite{4,5,6}. Surprise measures the probability of finding a particular number of intracommunity links in a given partition of a network, assuming that those links appeared randomly (see Methods for details and S formula). We have shown that S can be used to efficiently evaluate algorithms in open benchmarks and that, according to its results in those benchmarks, the best currently available algorithm turns out to be combining multiple methods to maximize S \cite{6}. These results suggest that Surprise may also contribute to evaluate algorithm performance in closed benchmarks and raise the question of whether S maximization could also be the best method to obtain optimal partitions in these complex benchmarks.

In this study, we carry out an extensive and detailed analysis of the behavior in closed benchmarks of a set of algorithms already used in open benchmarks in one of our recent papers \cite{6}. Our work has three well-defined sections. First, we test all those strategies in both LFR and RC closed benchmarks, being able to identify the algorithms which perform well and those that perform poorly or are unstable. Second, we propose a novel approach to compare methods, which involves hierarchically clustering all their solutions. Applying this procedure at different stages of the closed benchmarks, we obtain a better understanding of how the algorithms behave. Finally, we show that, as already demonstrated in open benchmarks, Surprise maximization is, among all tested, the best strategy for community characterization in closed ones.

\section{Results}

\begin{figure*}[ht]
\begin{center}
\includegraphics[scale=1]{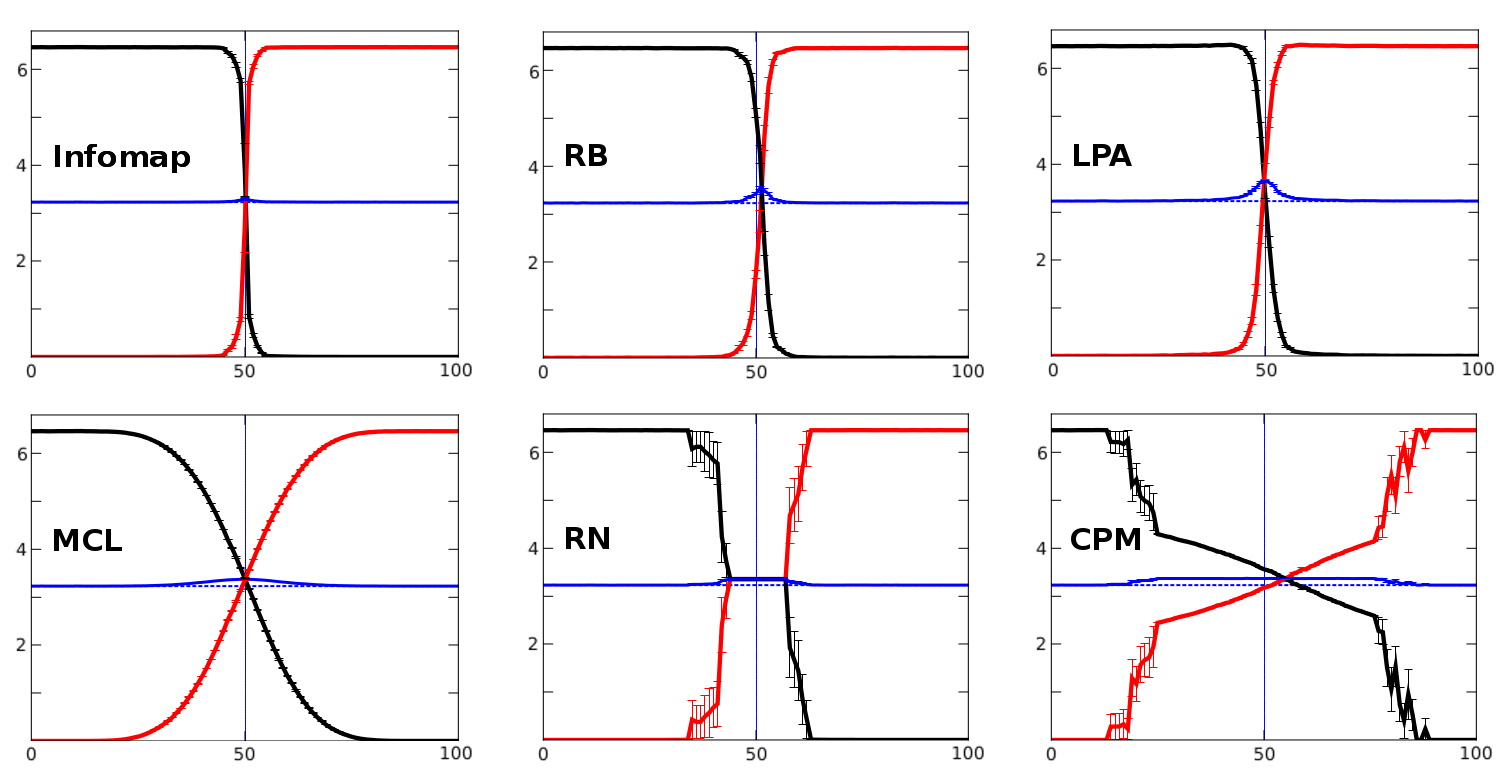}
\caption{\label{fig:1} Best algorithms in LFR closed benchmarks. The six algorithms able to recover the initial partition when C $\geq$ 5 \% are shown. In these diagrams, the x-axis shows the conversion percentage and the y-axis, the VI value. The red line indicates the VI values obtained when the algorithm solution is compared with the initial structure and the black line, the same comparison, but with the final structure. A perfect identity corresponds to the value VI = 0. Comparing the (VI$_{IE}$ + VI$_{EF}$)/2 values (blue line) and the VI$_{IF}$/2 values (dotted line, often invisible, being just below the blue one), we can conclude that Infomap, RB and LPA achieve optimal values until C is very close to 50 \%. MCL, RN and CPM work accurately only in the easiest analyses (both ends of the benchmark).}
\end{center}
\end{figure*}

\begin{figure*}[ht]
\begin{center}
\includegraphics[scale=1]{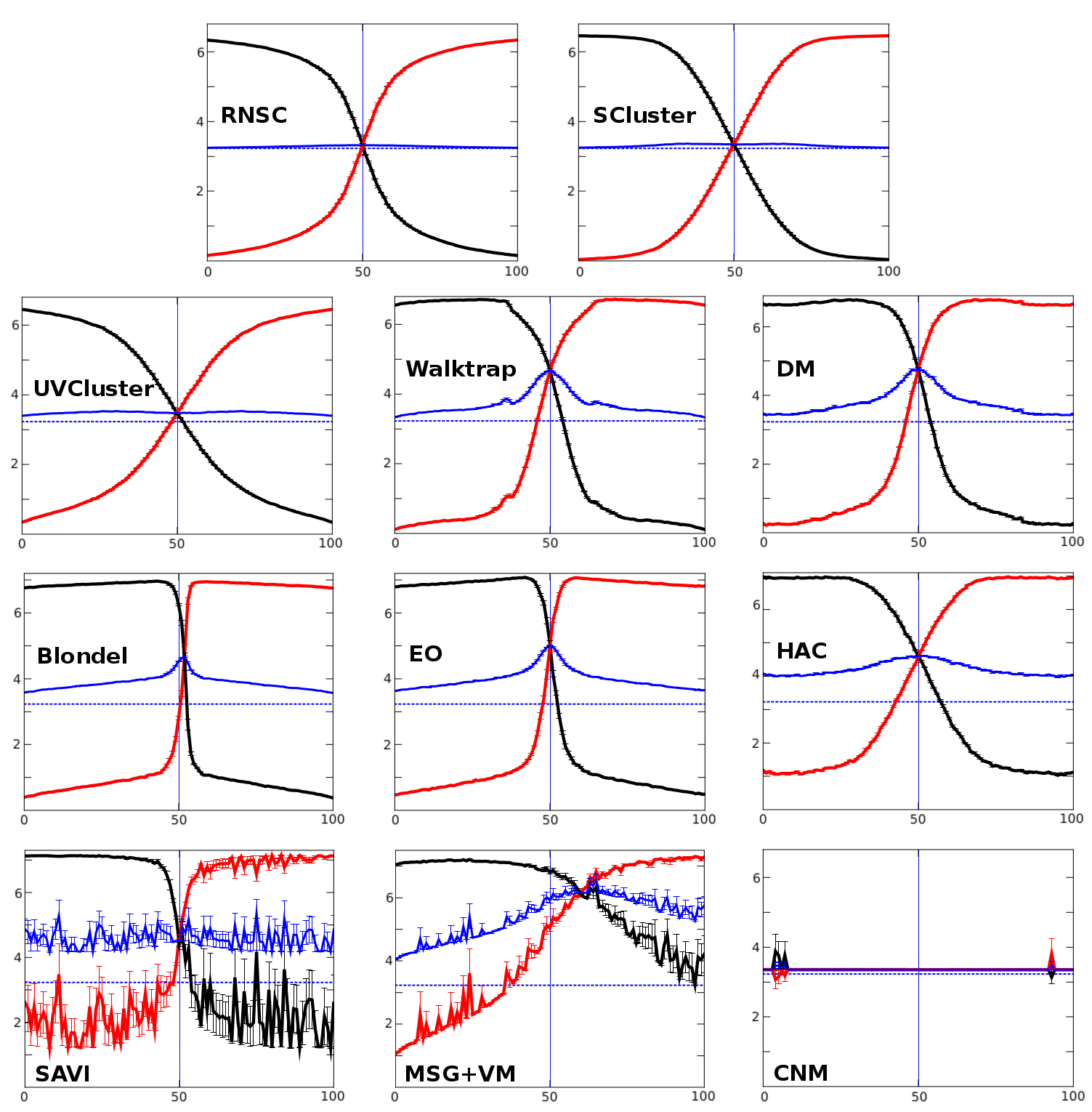}
\caption{\label{fig:2} Poor performers in LFR closed benchmarks. These algorithms were unable to recover, even once, the correct partitions of the benchmark. The plots show their very diverse behaviors, ranging from results resembling somewhat those shown in Figure \ref{fig:1} (RNSC or SCluster) to others that are highly asymmetric (MSG+VM), unstable (SAVI) or correspond to algorithms that fail to find any structure (CNM).}
\end{center}
\end{figure*}

\begin{figure*}[ht]
\begin{center}
\includegraphics[scale=1]{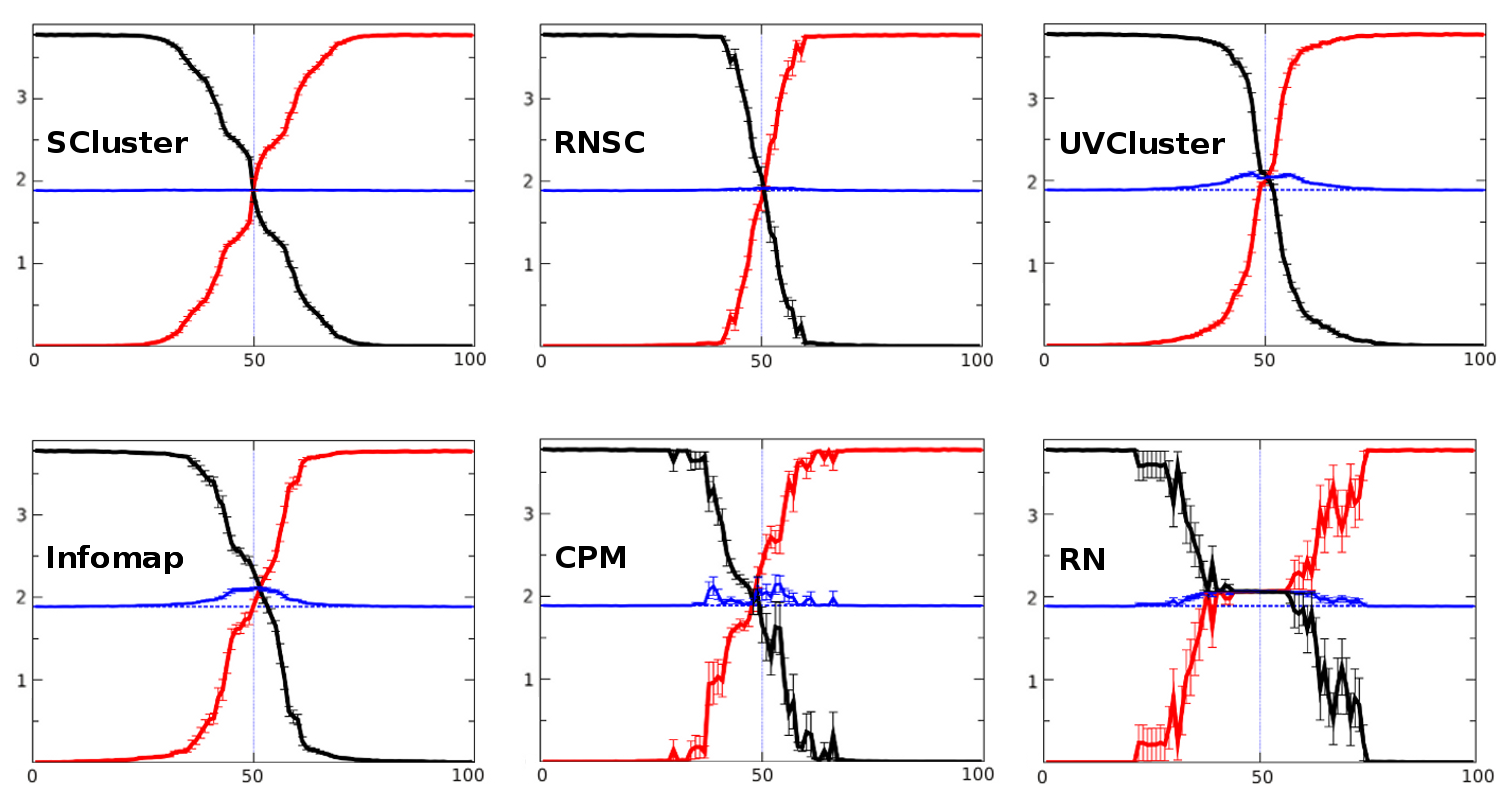}
\caption{\label{fig:3} Best algorithms in RC closed benchmarks. As in Figure 1, this figure shows the six algorithms that recovered the initial partition when C $\geq$ 5 \%. SCluster and RNSC showed an excellent behavior, displaying an almost straight blue line, while UVCluster failed in the central, most difficult, part of the benchmark. Infomap and CPM results were somewhat asymmetric, with the latter showing also some degree of instability. RN totally collapses when communities are not well defined.}
\end{center}
\end{figure*}

\begin{figure*}[ht]
\begin{center}
\includegraphics[scale=1]{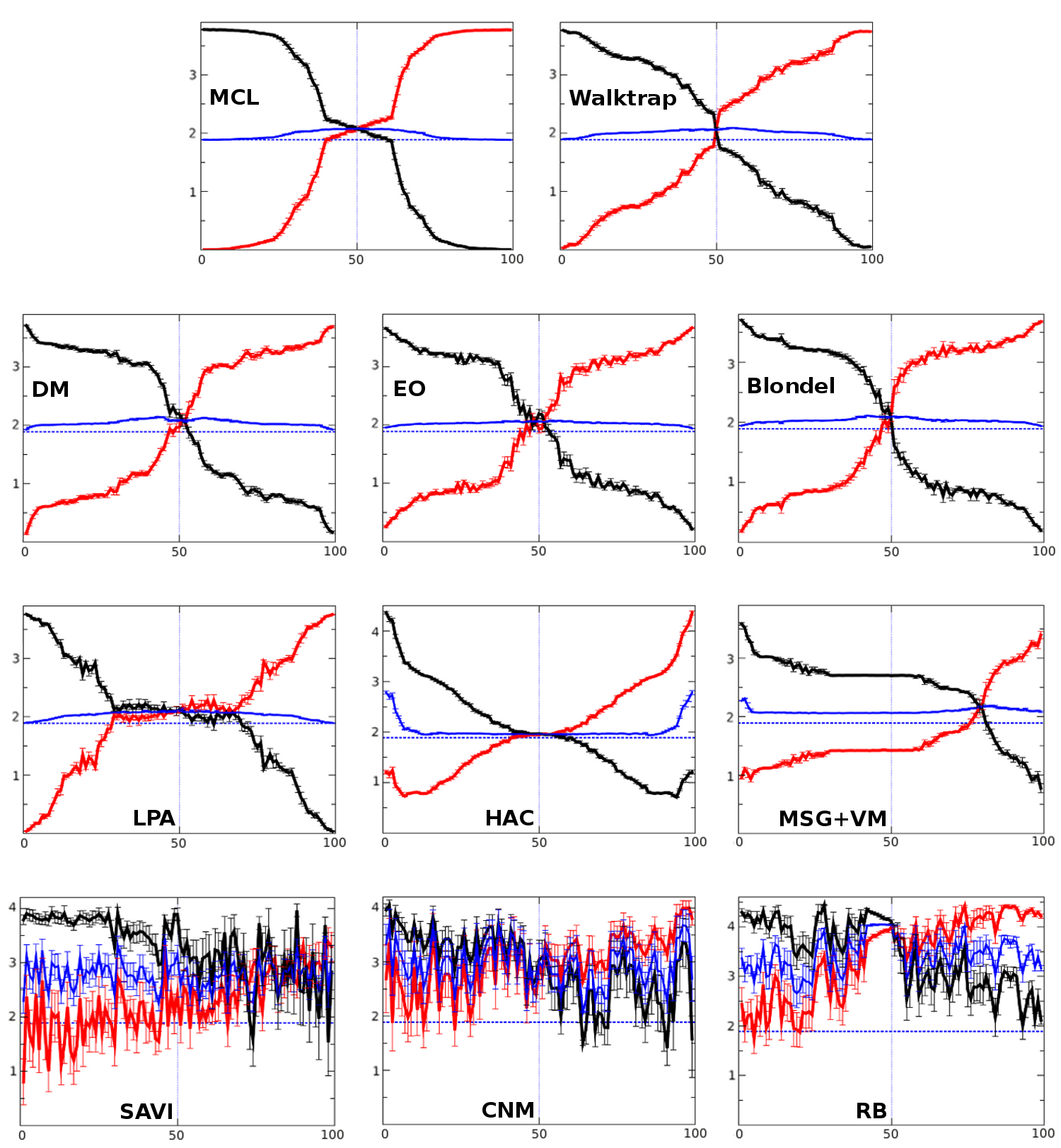}
\caption{\label{fig:4} Algorithms that performed poorly in RC closed benchmarks. In this case, the behavior of the algorithms was worse than in the LFR benchmarks showed in Figure \ref{fig:2}. MCL worked well only at the very beginning and the very end of the benchmark. The remaining algorithms performed much worse. In particular, MSG+VM showed a very asymmetric pattern and SAVI, CNM and RB results were chaotic.}
\end{center}
\end{figure*}

\subsection{Detailed behavior of the algorithms}
The 17 algorithms tested in the LFR and RC closed benchmarks showed very different behaviors, which are summarized in Figures 1 - 4. In these figures, following methods developed in previous works \cite{4, 5}, we show the VI values comparing the partitions obtained by the algorithms with the known initial (red lines) and final (black lines) structures. A perfect agreement with any of these structures corresponds to VI = 0. Also, the value (VI$_{IE} + $VI$_{EF}$)/2, (where E is the partition suggested by the algorithm, while I and F are, respectively, the initial and final partitions) is indicated with a blue line. As we discussed before, if the performance of an algorithm is optimal, then VI$_{IE}$ + VI$_{EF}$ = VI$_{IF}$. This means that, in these representations, the blue line should, in the best case, be perfectly straight and located just on top of a thin dotted line also included in these figures, which corresponds to the value VI$_{IF}$/2.

\begin{figure}[ht]
\begin{center}
\includegraphics[scale=.55]{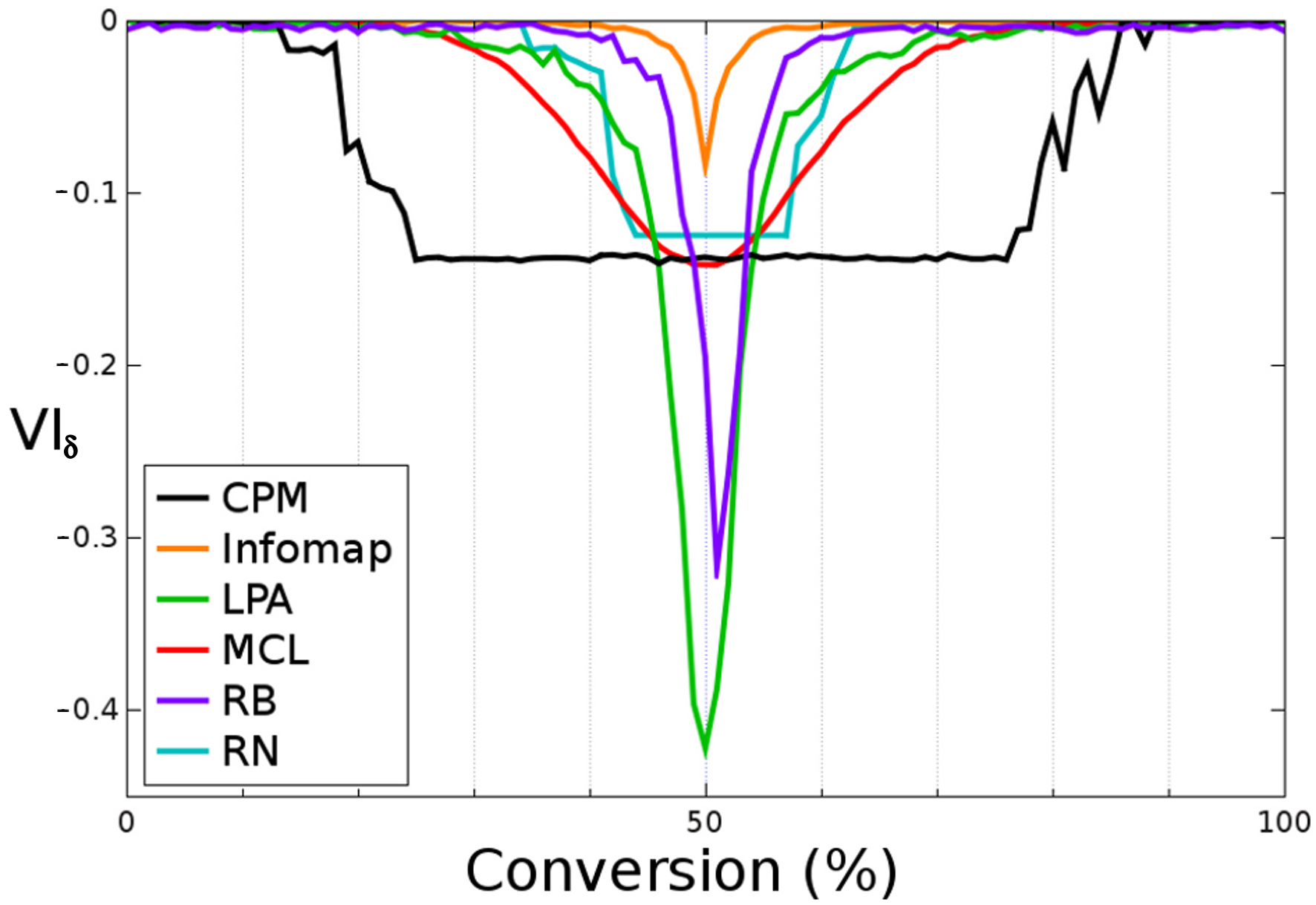}
\caption{\label{fig:5} Details of the performance of the best algorithms in LFR benchmarks. The y-axis (VI$_\delta$) corresponds to the difference between the expected value, VI$_{IF}$  and the VI$_{IE}$ + VI$_{EF}$ value of the different solutions. VI$_\delta$ values close to zero correspond to the best performers.}
\end{center}
\end{figure}

Figure 1 shows the behavior of the six algorithms in the LFR benchmarks that we considered the best, given that they were the only ones able to recover the initial partition when C $\geq$ 5 \%. Actually, none of the other 11 algorithms recovered even a single optimal partition in the whole benchmark. Given that C = 5 \% is certainly a very limited level of rewiring, these results indicate that most algorithms performed very deficiently. The six best algorithms worked however quite well, as indicated by the general closeness of their (VI$_{IE}$ + VI$_{EF}$)/2 values and the expected VI$_{IF}$/2 values (Figure 1).  Among these algorithms, Infomap \cite{26} was the only one able to perform optimally or quasi-optimally along the whole conversion process, although, around C = 50 \%, a slight deviation was noticeable (see blue line in Figure 1). Infomap recognizes the initial communities until almost half of the benchmark (red line with values VI  =  0) and then, just after C = 50 \%, it suddenly starts detecting the final ones (as seen by the fact that the black line quickly drops to zero). This rapid change is explained by the very similar sizes of all the communities present in the LFR benchmarks, which are all destroyed at the same time and also rebuilt all together with their final structure as conversion proceeds. Two other algorithms, RB \cite{30} and LPA \cite{27} performed quite similarly to Infomap, again only failing in the central part of the benchmark. The behavior of the other three among the six best-performing algorithms (MCL \cite{28}, RN \cite{31} and CPM \cite{20}), was good at the beginning of the conversion process, but clearly worse than Infomap quite soon (Figure 1). Figure 2 shows the results for the other algorithms. In addition of all them not finding any optimal solutions, the worst ones showed highly unstable solutions (e. g. SAVI \cite{33}, MSG+VM \cite{29}; notice the large mistakes in Figure 2) or totally collapsed, not finding any structure in these networks (e. g. CNM \cite{21}). We conclude that the behavior of some of the supposedly best algorithms, as defined by their performance in open benchmarks \cite{4,6,7,8,9,10,11,22,30,34}, is questionable when analyzed with more precision in these difficult closed benchmarks.

\begin{figure}[ht]
\begin{center}
\includegraphics[scale=.55]{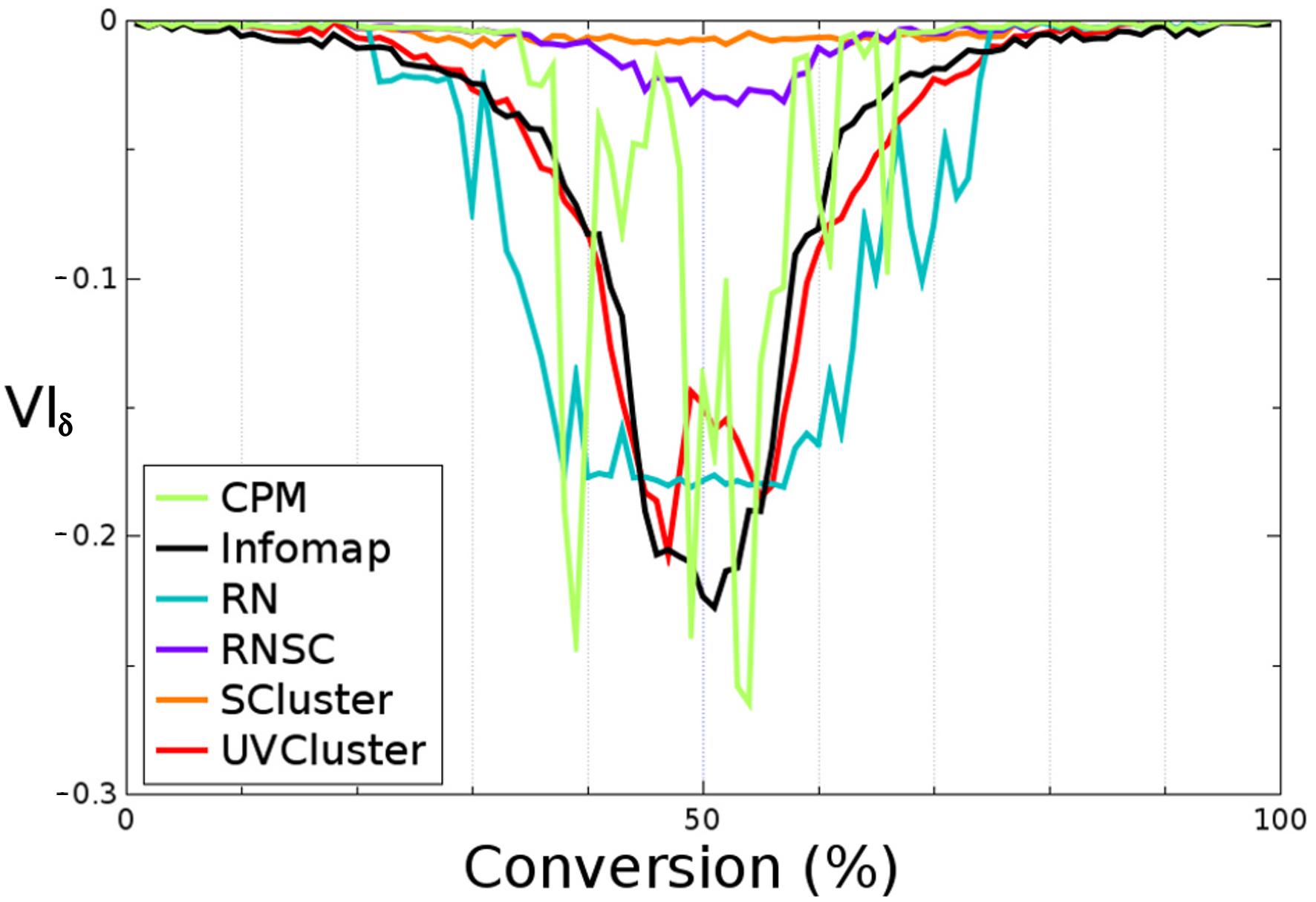}
\caption{\label{fig:6} Detailed performance in the RC benchmarks. Again, the better a performance, the closer to a value equal to zero.}
\end{center}
\end{figure}

In general, the results of the RC benchmark are similar. Again only six algorithms (Figure 3) provided correct values when C $\geq$ 5 \%. Interestingly, just three, Infomap, RN and CPM, passed the C $\geq$ 5 \% cut in both this benchmark and in the LFR benchmark (Figures 1 and 3). However, very significantly, none of these three were among the top performers in the RC benchmark. In the latter, we found that only SCluster \cite{34} and RNSC \cite{32} achieved optimal VI values along most of the conversion process (see again the blue lines in Figure 3). The remaining four algorithms that passed the C $\geq$ 5 \% cutoff (UVCluster \cite{33,34}, Infomap, CPM and RN) worked well during the easiest parts of the benchmark but failed when conversion approached 50 \%, in some cases showing asymmetries (CPM and Infomap) or instabilities (CPM and RN). These problems become much more noticeable in the worst algorithms, those that failed the 5 \% conversion cut (Figure 4). Again, the results for these algorithms are so poor that can be considered flawed. A final point is that, in the results provided by the best algorithms, a sudden swap from the initial to the final structure at around C = 50 \% does not occur, contrary to what we saw in the LFR benchmarks. This is due to the great variability in community sizes in the RC benchmarks, which causes them to disappear at different times of the conversion process.

Figures 5 and 6 show in more detail the deviations from the optimal values, indicated as VI$_\delta$ = VI$_{IF}$ - (VI$_{IE}$ + VI$_{EF}$), of the six best algorithms of each benchmark. This value is equal to 0 when agreement with the optimal performance is perfect. The larger the deviations from the optimal behavior, the more negative are the values of VI$\delta$. In the LFR benchmarks (Fig. 5), we confirmed that Infomap outperformed the other five algorithms. Its solutions were just very slightly different from the optimal ones around C = 50 \%. The other algorithms displayed two different types of behaviors. On one hand, MCL, RB and LPA progressively separated from the optimal value toward the center of the benchmark. Notice, however, that this minimum should appear exactly when C = 50 \%, this not being the case for RB, which showed slightly asymmetric results (Figure 5). On the other hand, RN and CPM reached a fixed minimum value that was maintained during a large part of the evolution of the network. This means that, during that period, these algorithms were constantly obtaining the same solution regardless of the network analyzed. In fact, RN allocated all nodes to different communities while CPM split the 5000 nodes into groups with one to four units. Figure 6 displays the analogous analyses for the RC benchmarks. We confirmed that SCluster and RNSC were clearly the best-performing strategies. The other algorithms satisfied the condition of optimality only when the network analyzed was very similar to either the initial or the final structure. This detailed analyses also showed more clearly something that could be suspected already looking at Figure 3, namely that RN and CPM produced abnormal patterns. The quasi-constant value of RN around C = 50 \% is explained by the fact that all its solutions in the center of the benchmark consisted of two clusters, one of them containing more than 99 \% of the nodes. On the other hand, CPM displayed an unstable behavior. The results in Figures 1-6 indicate that the RC benchmarks are at least as difficult as the LFR benchmarks, even though the number of nodes is much smaller (512 versus 5000). The considerable density of links and the highly skewed distribution of community sizes in the RC networks explain this fact.

\subsection{Hierarchical analysis of the solutions provided by the different algorithms}

\begin{figure*}[ht]
\begin{center}
\includegraphics[scale=1]{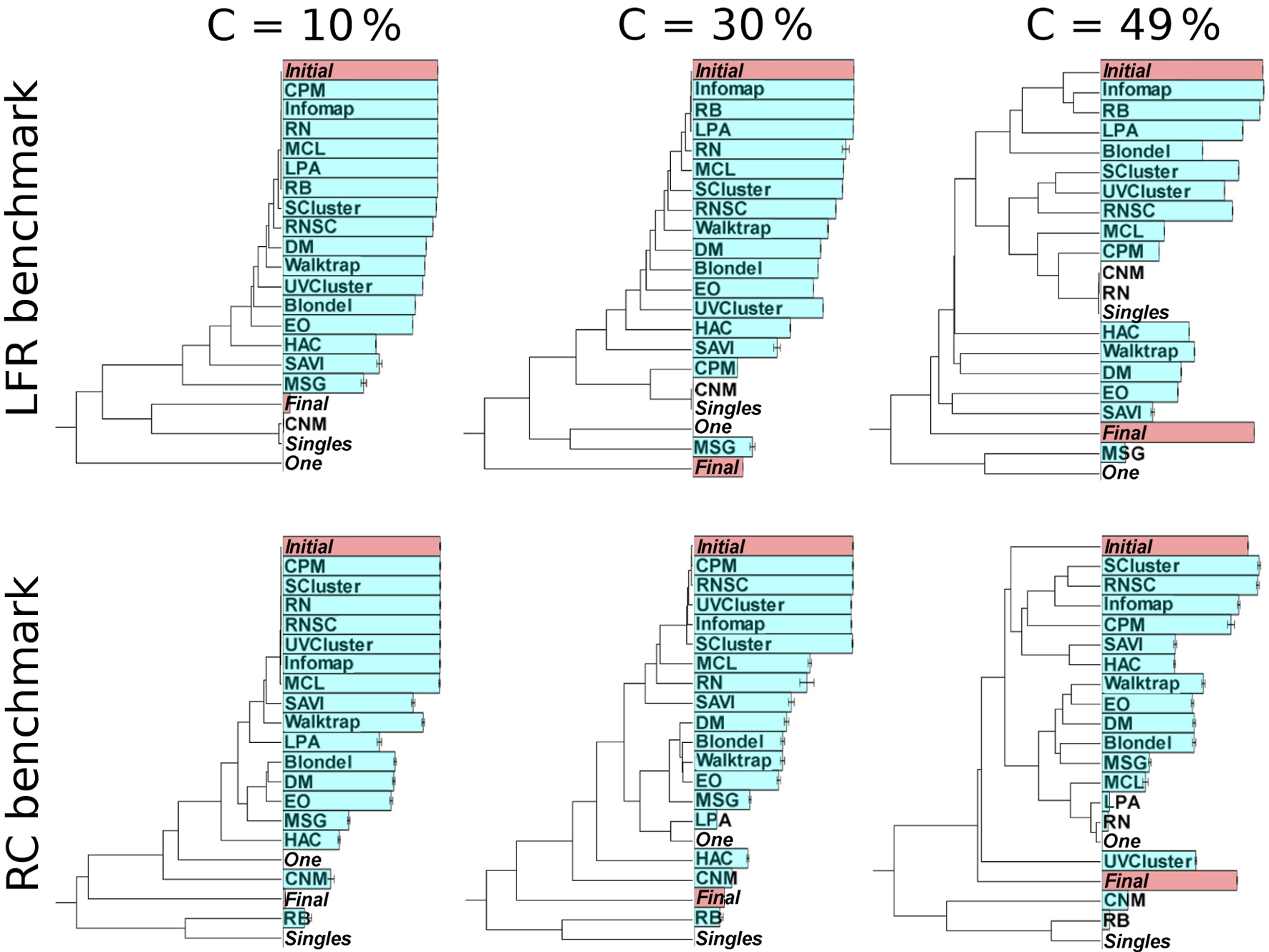}
\caption{\label{fig:7} Hierarchical clustering of solutions. Dendrograms representing the hierarchical clustering of the solutions achieved by the different methods in LFR (top panels) and RC (lower panels) closed benchmarks. Three different stages of the network conversion process have been analyzed: C = 10 \%, 30 \% and 49 \%. The four predefined structures (\textit{Initial, Final, One and Singles}) are indicated in italics.}
\end{center}
\end{figure*}

\begin{figure*}[ht]
\begin{center}
\includegraphics[scale=.9]{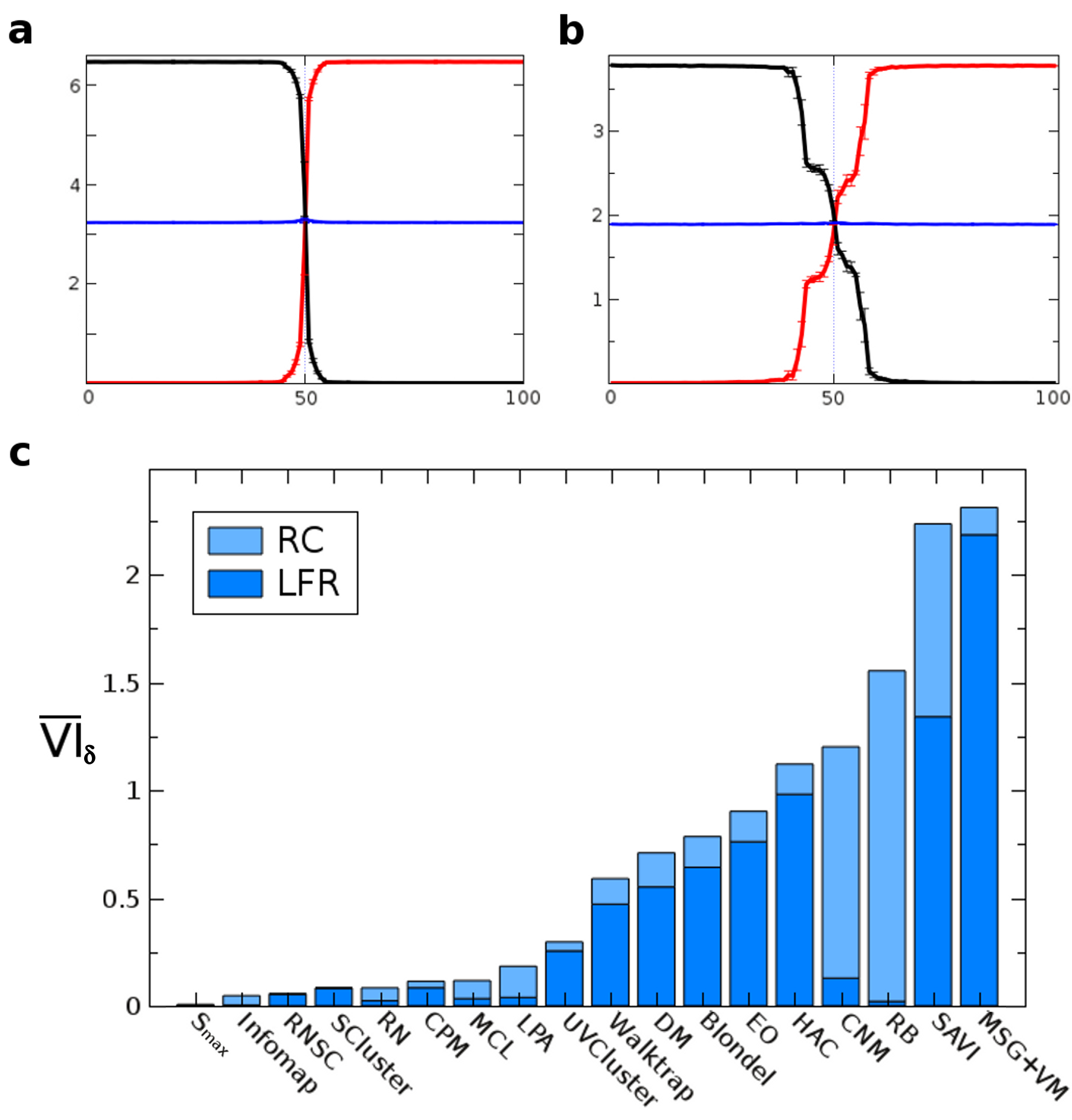}
\caption{\label{fig:8} Results of the $S_{max}$ meta-algorithm. Performance in LFR (panel a) and RC (panel b) benchmarks of the meta-algorithm that selected for each network the solution, among all the ones provided by the algorithms, which had the highest Surprise value. Panel c): Average values of the distance to the optimal performance (defined as the averages of the absolute values of VI$_\delta$) for all the algorithms.}
\end{center}
\end{figure*}

We considered it would be interesting to devise some way to understand at a glance the relationships among the solutions provided by multiple algorithms. Also, for closed benchmarks, that method should allow to visualize whether each algorithm is able or not to find good solutions when the community structure is being altered along the conversion process. Optimally, the method should be at the same time simple and quantitative. With these considerations in mind, we finally decided to perform hierarchical clusterings of the partitions generated by the algorithms at different C values, using VI as a distance. We considered also useful to include some additional partitions which would serve as reference points to better understand the results. 

Thus, as indicated in detail in the Methods section, we clustered the VI values of the solutions of all the algorithms, together with four artificial partitions (called \textit{Initial}, \textit{Final}, \textit{One} and \textit{Singles}), which respectively correspond to the initial and final structures present in the benchmark, a partition where all nodes are included in one community and a partition in which all nodes are in separated communities. These analyses were focused on three different stages of the benchmark, C = 10 \%, C = 30 \% and C = 49 \%. The first two were selected because they respectively corresponded to a low and medium degree of community structure degradation. We reasoned that any reasonable algorithm should easily recover the initial partition if C = 10 \%, while the results shown in the previous section indicated that, when C = 30 \%, the communities are fuzzier but still clearly detectable by several algorithms. Finally, when C = 49 \%, the initial communities should be in the limit of being substituted by the final ones. However, good solutions should still be slightly more similar to the initial partition than to the final one. Figure 7 displays the dendrograms for those three stages in both benchmarks, LFR and RC. Interestingly, these dendrograms, based on the matrix of VI distances, allow for a quantitative evaluation of how similar are the solutions provided by the different algorithms, given that the branch lengths are proportional to the corresponding distances of the matrix. We also include in that figure the Surprise values for each partition, as an independent measure of its quality (see below).

The LFR trees (Figure 7, top panels) display the behavior that could be expected after the detailed analyses shown in the previous section. Several of the best algorithms (e. g. Infomap, RB, LPA), appear in the tree very close to Initial even when C = 49 \%, showing that they are indeed recognizing the initial structure or very similar ones along the whole benchmark. However, it is clear that the distances from Initial to the solutions provided by the different algorithms are growing with increasing values of C. This indicates that the structures recognized by even the best algorithms are not exactly identical to the original ones, in good agreement with the results shown in Figure 5. In the case of the RC benchmark (Figure 7, bottom panels), the results are somewhat more complex. When C = 10 \% or C = 30 \%, the situation is very similar to the one just described for the LFR benchmarks: the best algorithms generate solutions that are very similar to Initial, just as expected. However, when C = 49 \% we found that the best algorithms in these benchmarks (SCluster, RNSC) generate solutions that are separated from Initial in the tree. Interestingly, their solutions cluster with those of other algorithms that also performed quite well in these benchmarks, such as Infomap or CPM. These results admit two explanations. The first one would be that all the algorithms have a similar flaw, which makes them find false structures. The second is that, when C = 49 \%, they are all recognizing a third type of structure, very different from Initial and Final, which is indeed the real one present in the networks. The first explanation is very unlikely given that these algorithms use totally unrelated strategies (Table 1). However, to accept the second one, we should have an independent confirmation that this may be the case.

Surprise values can be used to obtain such confirmation. In Figure 7, those values are also shown as horizontal bars with a size that is proportional to the S value obtained for each algorithm. As it can be easily seen in that figure, there is a strong correlation between the performance of an algorithm according to S values and its proximity to the Initial solution. This shows that S values are indeed indicating the quality of a partition with a high efficiency, as we already demonstrated in previous works [4-6]. Notice also that the S values for Initial and Final become more similar as the conversion progresses. This was expected, given that, at  C = 50 \%, the optimal partition should be exactly halfway between the initial and final community structures, and therefore, these values must then be identical. The fact that, in both the LFR and RC benchmarks with C = 49 \%, there are structures different from the initial one is indicated by the S values for the Initial partition not being the highest. The S value of the Infomap partition is statistically significantly higher (p = 0.0043; t test) than Initial in the LFR benchmarks with C = 49 \%. The same occurs in the RC benchmarks with C = 49 \%: both the SCluster and the RNSC partitions have Surprise values significantly higher than the one found for Initial (p $<$ 0.0001 in both cases; again, t tests were used). These results indicate that the top algorithms in these benchmarks are recognizing real, third-party structures, very different from Initial and Final, which emerged along the conversion process.

If the inclusion of Initial and Final was obviously critical for our purposes, the fact that we have also included One and Singles allows to visualize at a glance how some algorithms collapse, failing to find any significant structures in these networks. In the LFR benchmarks, this happens for CNM (already when C = 10 \%), RN, CPM and MSG+VM. All of them generate partitions very similar to either One or Singles. In the RC benchmarks, this same problem occurs with RB (again already with C = 10 \%), LPA, RN and CNM. We can conclude that these algorithms are often insensitive to the presence of community structure in a network. Notice that the combination of the VI$_\delta$-based analyses (Figures 5, 6) with these novel hierarchical analyses (Figure 7) allows establishing the performance of the algorithms with a level of detail and precision that is not currently attainable in open benchmarks.

\subsection{Surprise maximization results}
It is obvious from all the analyses shown so far that most algorithms performed poorly in these very difficult benchmarks. Even those that worked very well in one type of benchmark often had serious problems detecting the expected partitions in the other one. In a recent work \cite{6}, we showed in open benchmarks that a meta-algorithm based on choosing for each network the algorithm that generated the solution with the highest Surprise value worked better than any isolated algorithm and provided values that were almost optimal. Here, following that same strategy, we strikingly confirmed those results in closed benchmarks. Figures 8a and 8b show the behavior of choosing the maximal value of Surprise ($S_{max}$) in, respectively, the LFR and the RC benchmarks. $S_{max}$ values were obtained selecting solutions from six algorithms in the case of the LFR benchmark (ordered according to the number of times that they contribute to $S_{max}$, as follows: Infomap, RN, CPM, LPA, RB and MCL) and seven algorithms in the RC networks (i.e. CPM, RNSC, RN, SCluster, UVCluster, Infomap and MCL, ordered in the same way). All the other failed to provide any Smax values. As expected for a very good algorithm, the blue lines obtained for the Smax meta-algorithm are almost straight in both benchmarks (Figures 8a, 8b). If we measure the average distances to the dotted, optimal line, i.e. the average of VI$\delta$ for all conversion values, we found that it is minimal for the $S_{max}$ meta-algorithm, and just slightly different from zero (Figure 8c), being clearly better than the results of all algorithms taken independently (also shown in Figure 8c).

\section{Discussion}
We recently showed that closed benchmarks have advantages over the commonly used open benchmarks to characterize the quality of community structure algorithms \cite{5}. The main advantage is that the behavior of an algorithm can be more precisely understood by controlling the rewiring process, which leads to two testable predictions that any good algorithm must comply. The first is just a general, qualitative feature, namely the symmetry respect to the initial and final configurations along the conversion process. The second prediction is much more precise, being based on the fact that the relationship VI$_{IE}$ + VI$_{EF}$ = VI$_{IF}$ indicates optimal performance. These interesting properties of the closed benchmarks were already tested with a couple of algorithms in a previous work {5}. Here, we extended those analyses to obtain a general evaluation of all the best available community structure algorithms in two types of closed benchmarks. The general conclusions of this work are the following: 1) Closed benchmarks can be used to quantitatively classify algorithms according to their quality; 2) None of the algorithms works efficiently in all benchmarks; 3) Surprise, a global measure of quality of a partition into communities, may be used to improve our knowledge of algorithm behavior; and, 4) Surprise maximization behaves as the best strategy in closed benchmarks, as it does in open ones {6}. We will now discuss, in turn, these four conclusions.

We have shown that algorithms can be easily classified according to their performance in closed benchmarks based on different parameters. As just indicated above, two of them (VI$_{IE}$ + VI$_{EF}$ = VI$_{IF}$ relationship, expected symmetry of the results) were already described in our previous works. In addition to these two fundamental cues, additional parameters have been used for the first time in this work. Among them, we have first considered the ability of the algorithms to detect the initial community structure present in the networks when conversion starts growing. The critical value C $\geq$ 5 \% has been used as a cutoff value to select the best algorithms, given that those that do not recognize the original structure even when C is as low as 5 \%, are clearly poor performers. Another feature used here was VI$\delta$, the distance to the optimal VI value, which was used both to explore in detail the behavior of the algorithms along the conversion process (Figures 5 and 6) or, as an average, to rate them in a quantitative way (Figure 8). Finally, a novel strategy, based on hierarchically classifying the algorithms using the VIs among their partitions as distances, has been also proposed (Figure 7). We have shown that it allows to determine very detailed aspects of the behavior of the algorithms, such as establishing that, at high C values, some algorithms group together, all proposing related community structures, which are however very different from both the initial and final ones (Figure 7).

If we now consider our results respect to how the algorithms performed, we must be pessimistic. The first cutoff, optimal performance beyond C = 5 \%, allowed establishing that only two algorithms, Infomap and RN, are working well in both LFR and RC benchmarks (Figures 1, 3). Further analyses showed that others, such as RNSC, SCluster, CPM, MCL, LPA or UVCluster work quite well in average (Figure 8). However, they typically perform well in one of the benchmarks, but poorly in the other one (see Figures 1 - 4). Finally, a single algorithm, RB, works very well in the LFR benchmarks, but disastrously in the RC benchmarks (as becomes clear in the results shown in Figures 1 and 4 and quantitatively evaluated in Figure 8). This behavior is caused by the inability of this particular multiresolution algorithm to detect the communities of very different sizes present in the RC benchmarks \cite{14,15}. The other algorithms totally failed in both the LFR and the RC benchmarks (Figures 2, 4, 7). On top of their general lack of power to find the subtle structures present in these benchmarks when C increases, they often showed asymmetries, which we noticed were sometimes caused by a dependence of the results on the order in which the nodes were read by the programs (not shown).

Several papers have examined many of the algorithms used here in open GN, LFR and RC benchmarks. The general conclusions of those works can be summarized as follows: 1) As indicated already in the Introduction section, the GN benchmark is too easy, with most algorithms doing well \cite{7,8} while the LFR and RC benchmarks are much more difficult, with many algorithms working poorly \cite{5,6,8}. This means that tests on the GN benchmark should not be used to support that new algorithms perform well; 2) Infomap is the best algorithm for LFR benchmarks, with several others (RN, RB, LPA, SCluster) following quite closely \cite{6,8,9,10}; 3) However, SCluster, RNSC, CPM, UVCluster and RN are the best algorithms in RC benchmarks \cite{5,6}. Therefore, the agreement of the results in LFR and RC open benchmarks is far from complete; 4) All modularity maximizers behave poorly \cite{6,9,10}. These results are in general congruent with the ones obtained here in closed benchmarks, but some significant differences in the details have been observed. Comparing the results of the 17 algorithms analyzed here using closed benchmarks (Figure 8) with the performance of those same algorithms in open benchmarks that start with the same exact networks \cite{6}, we found that the top four average performers (Infomap, RN, RNSC and SCluster) were exactly the same in both types of benchmark. However, several algorithms (most clearly, RB and SAVI) performed much worse here. These poor performances of RB and SAVI were due to their unstable behavior in RC benchmarks (Figure 4).

In recent works, we have shown that Surprise (S) is an excellent global measure of the quality of a partition \cite{4,5,6}. In this work, we have taken advantage of that fact to improve our understanding of how algorithms behave. The combination of the hierarchical analyses described above with Surprise calculations have allowed to establish the presence of third-party community structures that the best algorithms find, and which are different from both the initial and final structures defined in the benchmarks (Figure 7). These differences are small in the LFR benchmark, in which the best algorithms, Infomap and RB, suggested community structures which are very similar to the initial one, even when C = 49 \% (Figure 7, top). They are however quite considerable in the RC benchmark, in which the best algorithms, SCluster and RNSC, plus several other among the best performers, appear together in a branch distant from the initial structure when C = 49 \% (Figure 7, bottom).

In previous works, we proposed that, given that Surprise is an excellent measure for the quality of a partition into communities, a good strategy for obtaining that partition would involve maximizing S. However, S-maximizing algorithms do not yet exist. So far, only UVCluster and SCluster use Surprise maximization as a tool to select the best partition among those found in the hierarchical structures that those algorithms generate \cite{34,35}, but the true $S_{max}$ partition is often not found with those strategies (as shown in refs. \cite{4,5,6} and this work). Given that we have not yet developed an $S_{max}$ algorithm, we decided to use a meta-algorithm that involves choosing among all the available algorithms, the one that produced the highest S value. This simple strategy was recently shown to outperform all known algorithms in open benchmarks \cite{6}. In this work, we have shown that the same occurs in closed benchmarks (Figure 8). Even more significant is the fact that, both in open and closed benchmarks, there is only a limited room for further improvement: by combining several algorithms using their S values as a guide, we obtain performances which are almost optimal (\cite{6} and Figure 7). The interest of generating S-maximizing algorithms, which could improve even on the combined strategy or meta-algorithm used so far in our works, is clear.

In summary, we have shown the advantages of these strategies and of using complex closed benchmarks for community structure characterization and the potential of Surprise-based analyses for complementing those tests. We have also shown that all tested algorithms, even the best ones, fail to some extent in these critical benchmarks and that a Surprise maximization meta-algorithm outperforms all them. The heuristic potential of these closed benchmarks is clear. They can be used in the future by anyone interested in checking the quality of an algorithm. A program to generate the conversion process typical of the closed benchmarks that can be applied to any network selected by the user is freely available at \href{https://github.com/raldecoa/ClosedBenchmarks}{https://github.com/raldecoa/ClosedBenchmarks}.

\section{Methods}

\begin{table*}
\begin{center}
    \begin{tabular}{|c|c|c|}
        \hline
        \textbf{Name of the Algorithm} & \textbf{Strategy used by the algorithm}              & \textbf{References} \\ \hline
        Blondel               & Multilevel modularity maximization          & \cite{20}       \\ 
        CNM                   & Greedy modularity maximization              & \cite{21}       \\ 
        CPM                   & Multiresolution Potts model                 & \cite{22}       \\ 
        DM                    & Spectral analysis + modularity maximization & \cite{23}       \\ 
        EO                    & Modularity maximization                     & \cite{24}       \\ 
        HAC                   & Maximum Likelihood                          & \cite{25}       \\ 
        Infomap               & Information compression                     & \cite{26}       \\ 
        LPA                   & Label propagation                           & \cite{27}       \\ 
        MLGC                  & Multilevel modularity maximization          & \cite{28}       \\ 
        MSG+VM                & Greedy modularity maximization + refinement & \cite{29}       \\ 
        RB                    & Multiresolution Potts model                 & \cite{30}       \\ 
        RN                    & Multiresolution Potts model                 & \cite{31}       \\
		RNSC				  & Neighborhood tabu search                    & \cite{32}		 \\
        SAVI      			  & Optimal prediction for random walks         & \cite{33}     \\
        SCluster  			  & Hierarchical Clustering + Surprise maximization & \cite{34}     \\ 
        UVCluster 		      & Hierarchical Clustering + Surprise maximization & \cite{34,35} \\ 
        Walktrap              & Random walks + modularity maximization          & \cite{36}     \\ 
        \hline
    \end{tabular}
\end{center}
\caption{\label{table:1}Details of the algorithms used in this study. A description of the strategies implemented by the algorithms and the corresponding references are indicated.}
\end{table*}
\subsection{Algorithms and benchmarks used in this study}
In this work, we evaluated 17 of the best performing community detection algorithms, selected according to recent studies [Table \ref{table:1}; \cite{4,5,6,7,8,9,10,22,30,34}]. These algorithms were the same used in a recent work \cite{6}, except that we discarded here one of the programs (implementing an algorithm called MLGC), given that it was unable to complete the analyses. In general, the default parameters of the algorithms were used. For the UVCluster and SCluster algorithms, we used UPGMA as hierarchical algorithm and Surprise as evaluation measure. RB and CPM have a tunable resolution parameter ($\gamma$) which defines the type of communities that they obtain. Since the optimal value for such parameter cannot be defined a priori in the absence of information about the community structure of the graph, we tested, for each network, a wide range of values of $\gamma$ and chose as solution the most stable partition. The RB approach is equivalent to the original definition of modularity when $\gamma$ = 1 \cite{30}, so we varied the parameter from 0 to as far as 5, ensuring a high coverage of the possible values of $\gamma$. In the case of the CPM algorithm, we used 0 $\leq \gamma \leq$ 1, which is the  range defined for unweighted networks \cite{22}.

Two very different types of networks were used as initial input for our closed benchmarks. The first were standard LFR networks containing 5000 nodes, which were divided into communities having between 10 and 50 nodes. The distribution of node degrees and community sizes were generated according to power laws with exponents -2 and -1, respectively. Since it was essential that the initial communities were well defined, we used a ``mixing parameter'' $\mu$ = 0.1. This value means that in the starting networks each node shared only 10 \% of its links with nodes in other communities \cite{16}. As already indicated, LFR communities are small and very numerous, but their sizes are very similar, which may be a limitation. Pielou’s index \cite{37} can be used to measure the variation of community sizes. This index, which takes a value of 1 for networks with equal-sized communities, was 0.98 in these LFR benchmarks. Thus, we decided to use a second type of benchmark with networks having a much more skewed distribution of community sizes. To this end, we used the Relaxed Caveman (RC) configuration. The networks used in our RC benchmarks contained 512 nodes, split into 16 communities. The Pielou's Index for the distribution of their sizes was 0.75, meaning that the differences in community sizes were very high, spanning two orders of magnitude.

In order to control the intrinsic variation of our analyses, ten different networks with the features defined above were generated as starting points both for the LFR and for the RC configurations. For these 20 different closed benchmarks, we obtained 99 intermediate points between the initial and the final partitions, generated using conversion values ranging from C = 1 \% to C = 99 \% We expected many different structures, with varied properties, to be produced along these complex conversion processes, thus allowing a thorough test of the community structure algorithms.

\subsection{Clustering of solutions}
We devised an approach for algorithm evaluation in closed benchmarks that allows to compare their solutions and to easily visualize their relationships. In this type of analysis, all the partitions provided by the different algorithms for a given network plus four additional predefined structures were considered. These four structures were: 1) \textit{Initial} and 2) \textit{Final}, which respectively correspond to the community structures present at the beginning and the end of the conversion process; 3) \textit{One}, which refers to a partition in which all nodes are in the same community; and, 4) \textit{Singles}, which corresponds to a partition in which all communities have a single node. 

The method used was the following: we choose three conversion values (10 \%, 30 \% and 49 \%) and we calculated the VI values obtained by comparing the partitions generated for a given network by all the algorithms to be tested plus the four predefined structures just indicated. To minimize the variance of the VI values, 100 different networks were analyzed for each conversion value. In this way, a matrix of VI values was obtained for each conversion level. Including the 4 preestablished structures, this matrix has ([$k$+4]*[$k$+3])/2 values, being $k$ the number of algorithms. The values of this VI matrix were then used as distances to perform agglomerative clustering using UPGMA \cite{38}. In this way, dendrograms that graphically depicted the relative relationships among all partitions were obtained. Given that we are using distances, how similar are the solutions of the different algorithms can be precisely evaluated, by considering both the topology of the tree and how long the branches in these dendrograms are. As described in the Results section, the four predefined structures were included to be used as landmarks to interpret the dendrograms generated.

\subsection{Surprise analyses}
The quality of a partition can be effectively evaluated by its Surprise (S) value \cite{6}. S is based on a cumulative hypergeometric distribution which, given a partition into communities of a network, computes its probability in a random network \cite{4,35}. Let $F$ be the maximum possible number of links in a network with $n$ links, and $M$ be the maximum possible number of intra-community links given that partition with $p$ intra-community links. Surprise is then calculated with the following formula \cite{4}:

\begin{equation}
S=\displaystyle-\log\sum\limits_{j=p}^{min(M,n)} \frac{\binom{M}{j}{\binom{F-M}{n-j}}}{\binom{F}{n}}
\end{equation}

The higher the S value, the more unlikely (or ``surprising'', hence the name of the parameter) is the observed distribution of intra- and intercommunity links, meaning that the communities obtained are maximally connected internally and also maximally isolated from each other.

\bibliography{CB_alg}

\end{document}